\documentstyle[prl,aps,epsf,epsfig,floats,preprint]{revtex}
\begin{document}
\draft
\tighten
%@twocolumnfalse\endcsname

\title{
\hfill{\small{DOE/ER/40762-214}}\\
\hfill{\small{UMD-PP-01-027}}\\[0.6cm]
Parity-Violating Pion-Nucleon Coupling $h_{\pi NN}^{(1)}$ \\
from $\pi^+$-Electroproton Production Near the Threshold}
\author{Jiunn-Wei Chen and Xiangdong Ji}
\address{Department of Physics, University of Maryland,
College Park, MD 20742-4111}
\maketitle
\begin{abstract}
We study the possibility of measuring parity-violating
pion-nucleon coupling $h_{\pi NN}^{(1)}$ 
from $\pi^+$-electroproton production in the threshold 
region. We calculate the electron-helicity asymmetry 
in terms of $h_{\pi NN}^{(1)}$ and the Z-boson
exchange between the electron and proton in 
leading-order heavy-baryon chiral perturbation 
theory. From the result, we demonstrate 
that the coupling can be determined to 
an accuracy of $\sim 2 \times 10^{-7}$ in an experiment 
with incident electron energy 250 MeV, 
luminosity $10^{38}$ sec$^{-1}$ cm$^{-2}$, and a running time of
$10^7$ sec. 
\end{abstract}
\medskip
%{PACS numbers: 11.30.Er, 11.30.Rd, 13.60.Le.} 

\vspace{1cm}

Weak interaction effects in hadron structure and 
reactions are usually masked by much stronger strong 
and electromagnetic interactions. Only under special 
circumstances can they be studied through high-precision 
measurements of observables vanishing in 
the absence of weak processes. The cross section 
asymmetry originating from a single longitudinal 
polarization is such an observable. 
At low energy, parity-violating 
hadronic interactions can be systematically classified in the 
framework of effective field theories \cite{KS,HBChPT,BKM}. 
In chiral power counting, the most important is the 
isovector P-odd pion-nucleon coupling $h_{\pi NN}^{(1)}$ 
which is responsible for 
the longest range part of the parity-violating $\Delta I=1$ 
$NN$ forces \cite{KS,DDH,AH}, the nucleon and nuclear anapole moments
\cite{ZE,HHM,MH,SR,vanKolck,ZPHM}. In
quantum chromodynamics (QCD), its 
value is believed to be dominated by the $s$-quark contribution 
generated from the neutral current interaction\cite{DSLS}. 
Clearly, $h_{\pi NN}^{(1)}$ is one of the most important observables 
that characterize the size of the nonleptonic weak 
interactions in hadron systems.

In the last two decades, serious attempts 
have been made to pin down $h_{\pi NN}^{(1)}$,
and as of to date, considerable uncertainty 
remains. Reviews of the experimental and theoretical
status at the different stages of investigation
can be found in \cite{AH,HH,Oers}.
In many-body systems, parity-violating effects 
can be enhanced significantly by important many-body 
correlations and have been detected experimentally.  
However, because of the underlying theoretical 
uncertainty, they cannot be translated easily 
into the corresponding results for the fundamental 
parameters. A large discrepancy exists on  
the extracted parity-violating coupling from the
gamma asymmetry $A_\gamma$ from the polarized 
$^{18}$F decay \cite{F18} and from the anapole 
moment measurement in $^{133}$Cs atom \cite{Cs133}. 
In few-body 
systems, the theory is under better control; 
but the parity-violating effects are generally 
small. The existing experimental results, and hence
the extracted parameters, have
very large error bars \cite{oldfew}. New high-precision
experiments currently under way are expected to 
overcome the problem of accuracy. 
These include $\overrightarrow{n}%
p\rightarrow d\gamma $ at LANSCE \cite{LANSCE}, 
$\overrightarrow{\gamma} d\rightarrow np$ at Jefferson Lab
(JLab)\cite{JLab}
(however, see Ref. \cite{Des} for a different opinion on the 
sensitivity of this channel to $h_{\pi NN}^{(1)}$), 
and the rotation of polarized neutrons in helium at NIST 
\cite{oldfew}. 
Parity-violating observables related to a single 
nucleon are free of nucleon-nucleon interactions and thus
have important advantages. In addition, heavy-baryon
chiral perturbation theory (HB$\chi$PT) provides a 
rigorous tool for making theoretical analyses. 
Recently, Bedaque and Savage, and Chen, Cohen and Kao
have studied parity-violating effects in Compton scattering 
with a polarized target \cite{BS} or a polarized beam \cite{CCK}. 
In Ref. \cite{CJ}, the present authors have studied extraction
of $h_{\pi NN}^{(1)}$ in the pion photoproton production, 
$\overrightarrow{\gamma }p \rightarrow n\pi ^{+}$ at the 
threshold region.
While all these processes are easy to analyze theoretically,
the experimental practicality requires 
them to have a large unpolarized cross section 
and large parity-violating asymmetries. In these regards,
the pion production is more attractive. 

Following our previous work mentioned above \cite{CJ},
we consider in this paper the measurement of
$h_{\pi NN}^{(1)}$ from $\pi^+$ production in low-energy 
electron-proton scattering. 
Since the momenta involved in the process
are constrained to be much smaller than the 
nucleon mass, a leading-order (LO) 
calculation in HB$\chi$PT is appropriate within 30\% uncertainty 
\cite{HBChPT,BKM}. 
Theoretical studies of the same process have been carried 
out previously by Li, Henley and Hwang and by Hammer and 
Drechsel by considering the predominant
Z boson exchange effects in the context of meson exchange 
models\cite{LHH,HD}. The goal of the present study 
is to use $ep\rightarrow e'\pi^+n$ to measure the
leading hadronic parity-violating coupling, and 
therefore we focus mainly on the threshold region 
where the effect of $h_{\pi NN}$ is important
and where HB$\chi$PT applies.

The process that we are interested in is
\begin{equation}
\vec{e}(k)
+p(P_{i})\rightarrow e(k') 
+ \pi ^{+}(k_\pi)+n(P_{f})\ ,
\end{equation}
where $k^{\mu }=\left(E_e,\vec{k}\right) $, 
$P_{i}^{\mu }$, $k^{\mu'
}=\left(E_e',\vec{k}'\right)$, $k^\mu_\pi = 
(\omega_\pi, \vec{k}_\pi)$, and $P_{f}^{\mu }$ are the 
laboratory four-momenta of the initial electron, proton, 
final electron, pion, and neutron, respectively. 
The interaction between the electron and the hadrons is 
treated in the Bonn approximation, and the four-momentum of the 
intermediate vector boson $\gamma$ or $Z$ is denoted as
$q= k-k'$. To justify the use of chiral 
perturbation theory, we require the four-momenta of the virtual 
vector boson and pion and three-momentum of the 
nucleon be small as compared to the nucleon mass 
scale $M_N$. 

%%%%%%%%%%%%%%%%%%%%fig 1%%%%%%%%%%%%%%%%%%%%%%%%%%%%%%%%
\begin{figure}[t]
\begin{center}
\epsfxsize=15cm%10cm
\centerline{\epsffile{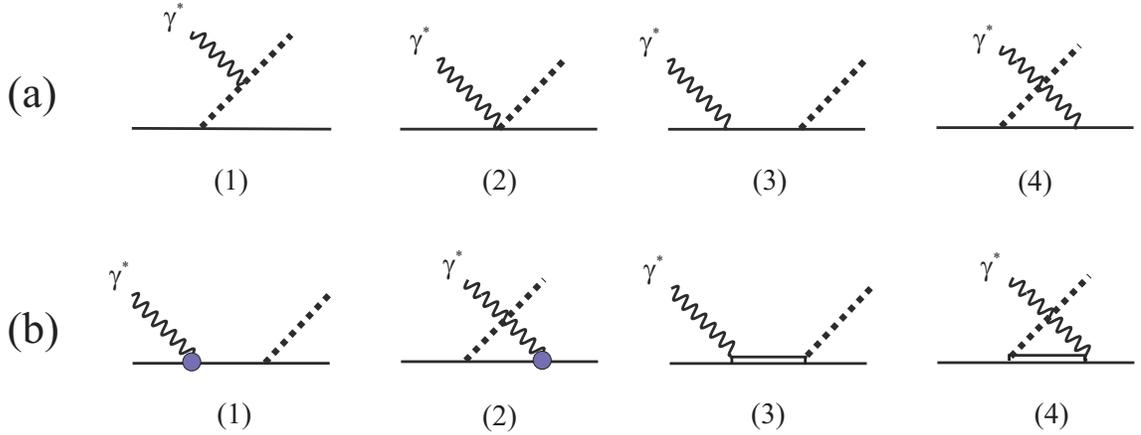}}
\end{center}
\caption{Leading order (a) and next-to-leading order (b) Feynman
diagrams for unpolarized  $\pi^+$ electroproduction from a proton.
The electron lines have been omitted and the delta resonance is 
represented by double lines. The solid circles denote 
the next-to-leading order vertices. The group (b) is incomplete 
but sufficient for the purpose of this paper.}
\label{fig:feypc}
\end{figure}

We first consider the unpolarized pion production 
cross section without weak interactions. In Fig. 1, 
we have shown the complete leading-order
(LO) (group (a)) and part of the next-to-leading order (NLO)
(group (b)) Feynman diagrams in HB$\chi$PT, where electron 
lines have been omitted, the wavy lines 
represent the virtual photons, solid lines the nucleons, 
dashed lines the pions, and the doule lines the 
$\Delta$ resonances. The solid circles denote the 
NLO vertices. The differential cross section after integrating
over the final neutron momentum can be written as 
\begin{equation}
     {d\sigma \over dE'd\Omega'}
      = {2\alpha_{\rm em}^2\over Q^4}
       {E'M_N\over E} ~{\cal M}~\delta
            \left((P_i+q-k_\pi)^2-M_N^2\right)~
         {d^3\vec{k}_\pi\over (2\pi)^3 2\omega_\pi} \ , 
\end{equation}
where $Q^2=-q^2$ and 
\begin{eqnarray}
     {\cal M}&=&\ell^{\mu\nu}~W_{\mu\nu}\ , \nonumber \\
      W_{\mu\nu} &=& \left\langle
             p(P_i)|J_\mu|n(P_f)\pi(k_\pi)\right\rangle
               \left\langle n(P_f)\pi(k_\pi)|J_\nu|p(P_i)
             \right\rangle \ , 
\end{eqnarray}
with $P_f =P_i+q-k_\pi$.
The unpolarized lepton tensor is $\ell^{\mu\nu}
= (1/2){\rm Tr}[\not\!k\gamma^\mu\not\! k'\gamma^\nu]
=2(k^\mu {k'}^\nu+k^\nu {k'}^\mu-g^{\mu\nu}k'\cdot k)$. 
The hadron tensor $W^{\mu\nu}$ depends on the matrix 
elements of the electromagnetic current between 
the hadron states. 
From the LO diagrams, we find 
\begin{eqnarray}
   W^{\mu\nu} &=& {g_A^2\over 2f_\pi^2}~\left[
             (2k_\pi-q)^\mu(2k_\pi-q)^\nu
              {|\vec{k}_\pi-\vec{q}|^2 \over ((k_\pi-q)^2-m_\pi^2)^2}
         \right.   \nonumber \\              
  && + {\vec{k}_\pi\cdot (\vec{k}_\pi-\vec{q})
              \over \omega ((k_\pi-q)^2-m_\pi^2)}
            \left[v^\mu(2k_\pi-q)^\nu + v^\nu(2k_\pi-q)^\mu\right]
                   \nonumber \\
          &&  -{1\over (k_\pi-q)^2-m_\pi^2}
     [\left((2k_\pi-q)^\mu v^\nu +(2k_\pi-q)^\nu 
         v^\mu\right) v\cdot(k_\pi-q) \nonumber \\
        &&  - (2k_\pi-q)^\mu(k_\pi-q)^\nu - 
            (2k_\pi-q)^\nu(k_\pi-q)^\mu ]
           \nonumber \\
          &&\left. \left(1 + {|\vec{k}_\pi|^2\over \omega^2}
               - {2\omega_\pi\over \omega}\right)v^\mu v^\nu
                - g^{\mu\nu}
        + {1\over \omega}
          \left(v^\mu k_\pi^\nu - v^\nu k_\pi^\mu\right) \right] \ , 
         \end{eqnarray}
where $v^\mu = (1,0,0,0)$ is the velocity of the nucleon 
at rest, $g_A=1.257$ is the neutron-decay constant,
and $f_\pi=93$ MeV is the pion decay constant. As expected
$W^{\mu\nu}$ is symmetric in $\mu$ and $\nu$, and satisfies
the current conservation: $W^{\mu\nu}q_\mu=0$. 

%%%%%%%%%%%%%%%%%%% Figure 2 %%%%%%%%%%%%%%%%%%%%%%%%
\begin{figure}[t]
\begin{center}
\epsfxsize=15cm%5cm
\centerline{\epsffile{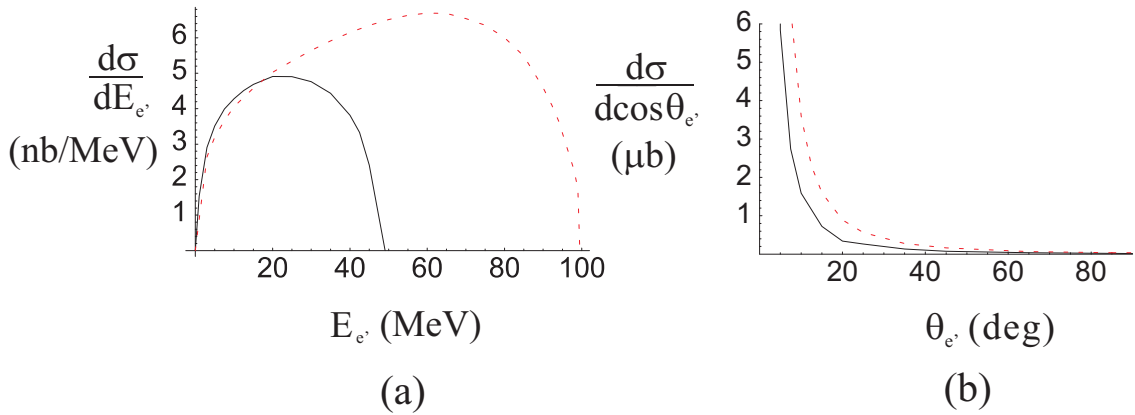}}
\end{center}
\caption{Pion-production cross section
as a function of the scattered electron energy (a) 
and polar angle (b) at the incident electron energy 
$E_e = 200$ (solid line) and $250$ MeV (dashed line), 
respectively.}
\label{fig:pipc}
\end{figure}

For the process to be useful in probing small  
parity-violating effects, the cross section must 
be sizable enough to yield a large number of 
events ($\sim 10^{14}$) in a realistic experiment.  
For an estimate, it is sufficient to examine the 
LO theoretical prediction. In Fig. 2(a), we 
have shown the cross section as a function of the 
scattered electron energy after all other variables
are integrated. The solid curve is the result with the incident 
electron energy $E_e=200$ MeV, and the dashed one with
$E_e=250$ MeV. In the former case,
the scattered electron energy is peaked around 20 MeV, 
and in the latter, it ranges from 20 to 80 MeV. In principle,
the electrons from elastic scattering off the protons 
have much higher energy. When 
taking into account quantum electrodynamics bremsstrahlung,
however, the elastic peak has a well-known tail which
contaminates the electron spectrum from  
pion production. 
In Fig. 2(b), we have shown the cross section as a function 
of the scattered electron angle. As expected, 
the cross section is forward peaked due to 
the infrared factor $1/Q^4$ in Eq. (2). For both beam 
energies, most electrons are scattered into 
a polar angle less than $10^\circ$.
 
%%%%%%%%%%%%%%%%%%%%%% Fig. 3 %%%%%%%%%%%%%%%%%%%%%%%%%%%
\begin{figure}[t]
\begin{center}
\epsfxsize=15cm%5cm
\centerline{\epsffile{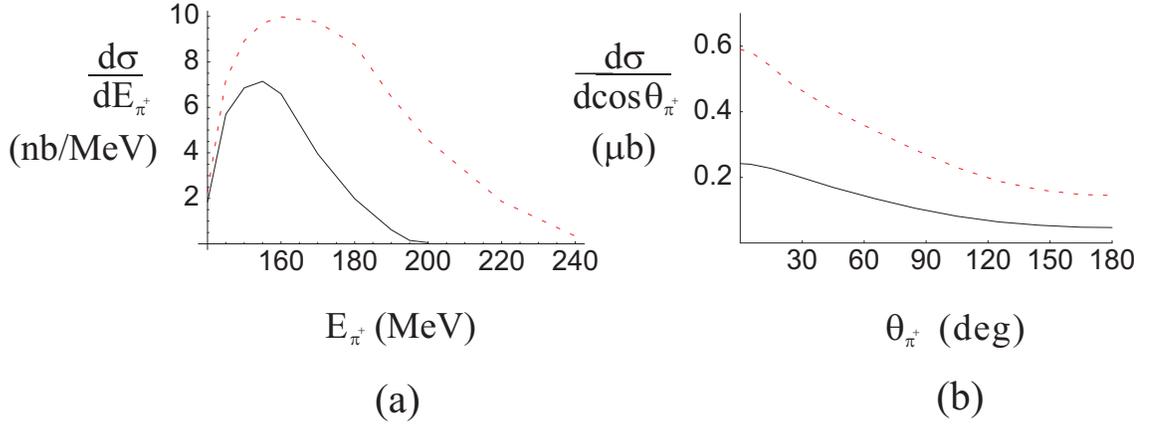}}
\end{center}
\caption{Pion-production cross section
as a function of the outgoing pion energy (a)  
and polar angle (b) at the incident electron energy 
$E_e = 200$ (solid line) and $250$ MeV (dashed line), 
respectively.}
\label{fig:epc}
\end{figure}

Because of the large number of events needed 
in parity-violating experiments, coincidence detection 
is not an option. For $ep\rightarrow e'\pi^+n$, 
the best possibility may be $\pi^+$ semi-inclusive detection.
In Fig. 3(a), we have shown the integrated 
cross section as a function of the energy of 
the pion yield. At $E_e=200$ MeV, 
the pion energy is peaked around $\omega_\pi=155$ MeV, which
translates to a pion momentum $\sim 65$ MeV, or 
$\beta=v/c\sim 0.45$. Therefore, the pion decay length 
is about 3.5 m, and most pions can be detected 
before their weak decays. 
At $E_e=250$ MeV, the pion energy is peaked around 170 MeV. 
In Fig. 3(b), we have shown
the pion angular distribution in lab frame. 
At backward angles, the production rate is about a factor of 
3 smaller than that at the forward angles. This reduction 
is mainly due to the Lorentz boost.
According to the figure, the semi-inclusive 
pion production cross section is about 100nb at a 
forward steradian and at $E_e=250$ MeV. 
To produce $10^7$ events per second needed in a useful
and practical experiment, the luminosity must be at least
$10^{38}$ sec$^{-1}\cdot$ cm$^{-2}$. Although the lower
limit is not difficult to achieve, a high-statistics 
measurement may require much higher luminosity, making 
the experiment challenging. 

%%%%%%%%%%%%%%%%%%%%%%%Fig 4.%%%%%%%%%%%%%%%%%%%%%%%%%%%%%%%%%%%%%%%%%%%
\begin{figure}[t]
\begin{center}
\epsfxsize=15cm%10cm
\centerline{\epsffile{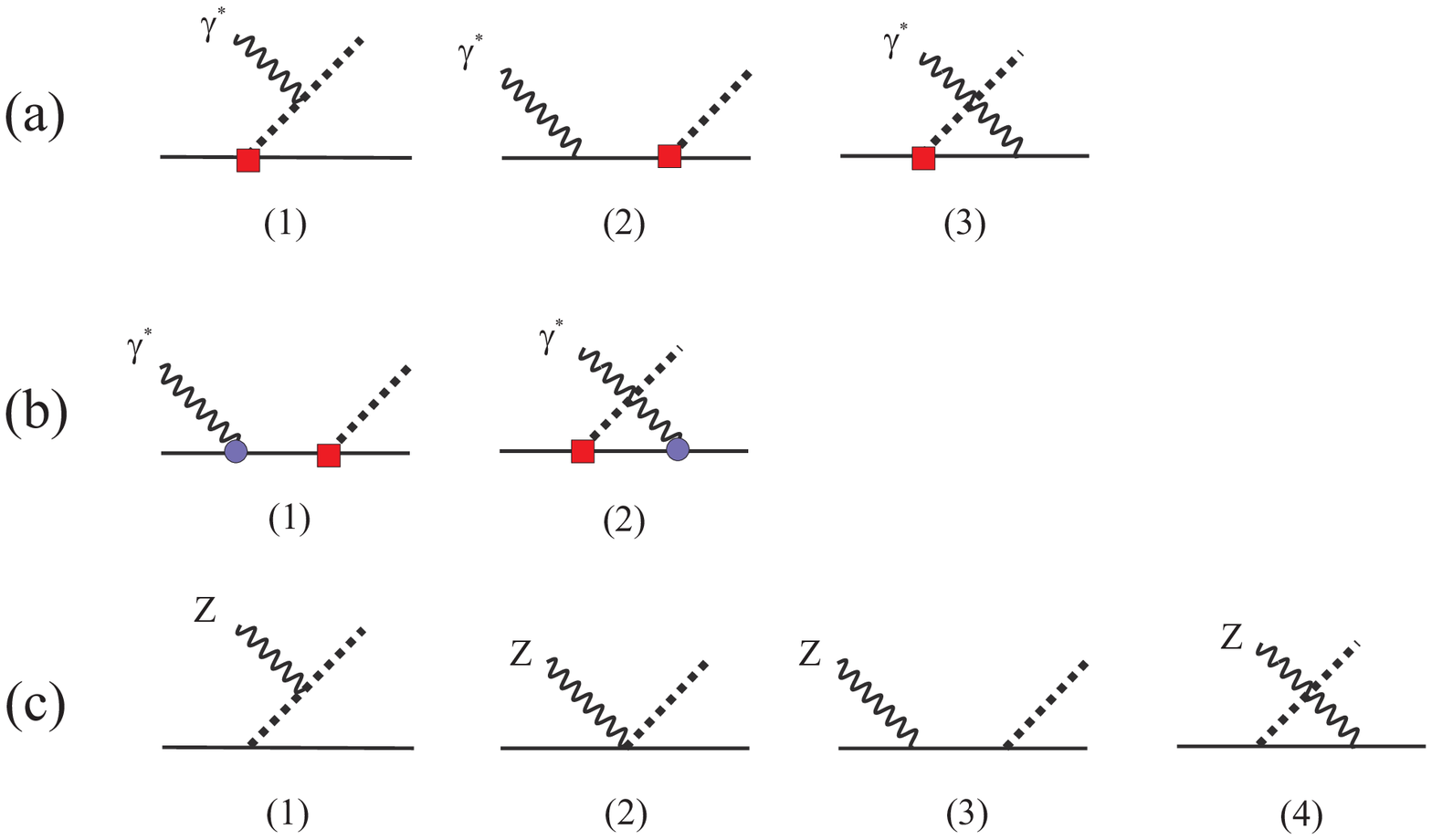}}
\end{center}
\caption{
Feynman diagrams contributing to the parity-violating amplitudes at
LO and NLO in 
$\protect\overrightarrow{e} p 
\protect\rightarrow e'\pi^{+}n$. }
\label{PVLO}
\end{figure}

Given the sizable rate for $\pi^+$ production in 
$e$-$p$ scattering, we now analyze the 
parity-violating effects. According to the standard 
electroweak theory, parity violation in $\vec{e}p\rightarrow
e'\pi^+n$ comes from two distinct sources. First, 
the non-leptonic weak interactions generate parity 
violation hadronic vertices. At low energy, 
these vertices can be classified systematically 
in terms of chiral power counting \cite{KS}. 
To ${\cal O}(p^0)$ (ignore the weak coupling 
in power counting), it has one term, 
\begin{equation}
{\cal L}^{PV}=-ih_{\pi NN}^{(1)}\pi ^{+}p^{\dagger }n
 +h.c.+\cdots \ , \label{Lw}
\end{equation}
where the ellipses denote terms 
with more pion fields and derivatives, and
the phase convention is taken from Refs.\cite{vanKolck,ZPHM}. 
The complete leading (group (a)) and part of the next-to-leading 
(group (b)) order Feynman diagrams involving $h_{\pi NN}^{(1)}$ 
are shown in Fig. 4. The contribution involving
the intermediate delta resonance comes in at higher
orders. The second source of parity-violation 
is the semi-leptonic process of Z-boson exchange
between the scattering electron and the nucleon system.
The relevant Feynman diagrams are shown in group (c) of 
Fig. 4. Since $\sin^2\theta_W \sim 1/4$, we approximate
the Z-electron coupling as pure axial, and the 
Z-quark coupling pure vector. 

The leading hadronic parity-violating
contribution comes from the interference of the LO
parity-conserving and LO parity-violating amplitudes; 
the result depends on the nucleon polarization.
For polarized electron scattering on unpolarized proton, 
the lepton tensor is $l_{\mu \nu}=2 i
\epsilon^{\mu \nu \alpha \beta}k_{\alpha}k'_{\beta}$ for helicity +
electron; the leading 
P-odd contribution comes from the interference between
the LO (NLO) parity-conserving and NLO
(LO) parity-violating amplitudes. Although the NLO diagrams
in Figs. 1(b) and 4(b) are incomplete, they are sufficient
for calculating this P-odd contribution.
The resulting hadron tensor $W^{\mu\nu}$ is 
\begin{eqnarray}
   W^{\mu\nu} &=& -{i g_A h_{\pi NN}^{(1)}\over \sqrt{2}
           f_\pi M_N \omega}\left\{
          \epsilon^{\mu\nu\alpha\beta} q_\alpha v_\beta
          \left(\mu_p-{\omega\over\omega_\pi} \mu_n\right) 
          \right. \nonumber \\
 && + \left[\epsilon^{\mu\nu\alpha\beta}q_\alpha v_\beta
          \left({2m_\pi^2-q\cdot k_\pi \over
                    (k_\pi-q)^2-m_\pi^2} + {\omega_\pi\over 
                  \omega}\right) \right. \nonumber \\
      &&  \left. + \epsilon^{\mu\nu\alpha\beta}k_{\pi\alpha}
               q_\beta\left(
                {2\omega_\pi-\omega 
                \over (k_\pi-q)^2-m_\pi^2} + {1\over \omega}
              \right)\right] \nonumber \\
        &&  \left. \times\left[-(\mu_p-{\omega\over \omega_\pi}
               \mu_n) + {2g_{\pi N\Delta}G_1\over 9g_A}
              \left( {\omega \over \omega -\Delta}
                + {\omega \over \omega_\pi + \Delta}
               \right)\right]\right\} \ ,  
\end{eqnarray}
which is conserved and antisymmetric 
in $\mu$ and $\nu$. Parameter $\Delta$ is the mass difference
between the nucleon and delta resonance, and is treated as small 
as compared to $M_N$. $\mu_{p,n}$ is the magnetic moment 
of the proton and neutron in units of the nuclear magnetons, 
respectively. $G_1=3.85$ and  $g_{\pi N \Delta}=1.05$ are the
$\gamma$-$N$-$\Delta$ and the $\pi$-$N$-$\Delta$ coupling constants 
\cite{Hemmert2}. 

As mentioned above, the semi-leptonic parity-violating 
contribution comes from the Z-exchange between the scattering
electron and the nucleon system. Taking into account only 
the axial coupling from the electron, the parity-violating 
S-matrix element is
\begin{equation}
    \langle e'\pi^+n|S|ep\rangle =
      - i(2\pi)^4\delta^4(k+P_i-k'-k_\pi-P_f)
      {G_F\over \sqrt{2}} \bar e(k')\gamma^\mu\gamma_5 e(k)
      \langle \pi^+ n|J^Z_\mu|p\rangle \ , 
\end{equation} 
where $J^{Z\mu}$ is the vector part of the quark neutral
current: $\bar u\gamma^\mu u(1/2-4/3\sin^2\theta_W)
   + \bar d\gamma^\mu d(-1/2 + 2/3\sin^2\theta_W) +...$. 
The LO matrix element $\langle \pi^+ n|J^Z_\mu|p\rangle$
in HB$\chi$PT can be calculated from Feynman diagrams
shown in Fig. 4(c). After approximating 
$\omega=\omega_\pi$, the resulting parity-violating 
cross section is directly proportional to the LO 
spin-independent cross section, 
\begin{equation}
      d\sigma^{\rm PV~ from ~Z} = 
       - 2\sqrt{2} (1-2\sin^2\theta_W)
          {G_FQ^2 \over 4\pi\alpha_{\rm em}} 
d\sigma^{\rm PC~ at~LO} \ . 
\end{equation}
One can calculate the Z-exchange more accurately
by including the electron vector-current contribution 
and going to NLO in HB$\chi$PT. For our purpose, however,
the above accuracy is sufficient. 

%%%%%%%%%%%%%%%%%%% Figure 5 %%%%%%%%%%%%%%%%%%%%%%%%
\begin{figure}[t]
\begin{center}
\epsfxsize=15cm%5cm
\centerline{\epsffile{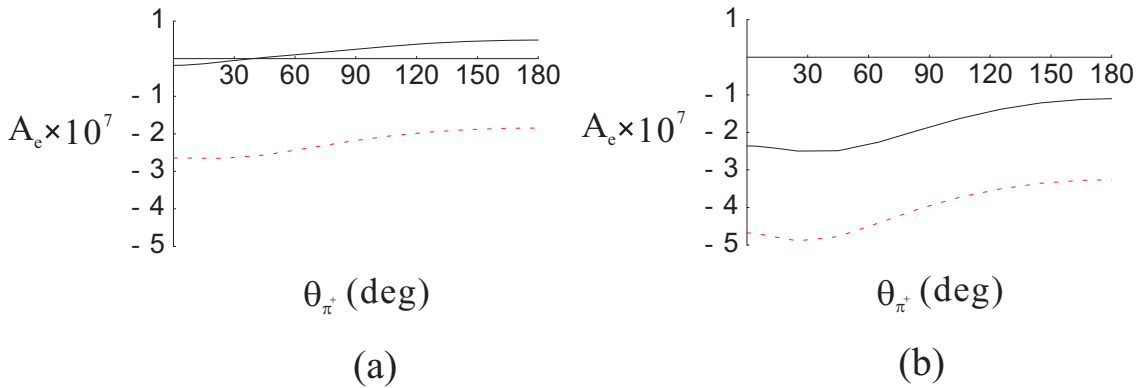}}
\end{center}
\caption{
The electron helicity asymmetry $A_e$ as a function
of outgoing-pion polar angle at $E_e=200$ (a) and
$250$ MeV (b). The dashed lines include the Z
contributions only, whereas the solid lines include
the contribution from $h_{\pi NN}^{(1)}$ (taken to be $5 \times
10^{-7}$) as well.}
\label{fig:asym}
\end{figure}

In Fig. 5(a), we have shown the electron-helicity asymmetry 
(defined as the difference between the positive
and negative helicity cross sections normalized to the sum) 
as a function of the polar angle of the pion 
in the laboratory frame. For $E_e=200$ MeV, the asymmetry
from Z-exchange (dashed line) 
is on the order of $-2.5\times 10^{-7}$
with small angular dependence. Assuming the ``best
guess" for $h_{\pi NN}^{(1)}=5\times 10^{-7}$ \cite{DDH}, 
the hadronic parity-violating
contribution cancels this almost entirely, as shown 
by the solid curve. The cancelation shows 
that, at this low energy, the two contributions have
the same order of magnitude. As a result, the total 
parity-violating asymmetry is difficult to measure.  
When the scattering electron energy is increased to 250 
MeV, the asymmetry from Z-exchange is increased by 
approximately a factor of 2 because of the larger $Q^2$, 
whereas the contribution from 
$h_{\pi NN}^{(1)}$ changes little, as shown in Fig. 5(b). 
The combined asymmetry is now negative, and has 
a sizable magnitude around $-2\times 10^{-7}$.   

%%%%%%%%%%%%%%%%%%% Figure 6 %%%%%%%%%%%%%%%%%%%%%%%%
\begin{figure}[t]
\begin{center}
\epsfxsize=15cm%5cm
\centerline{\epsffile{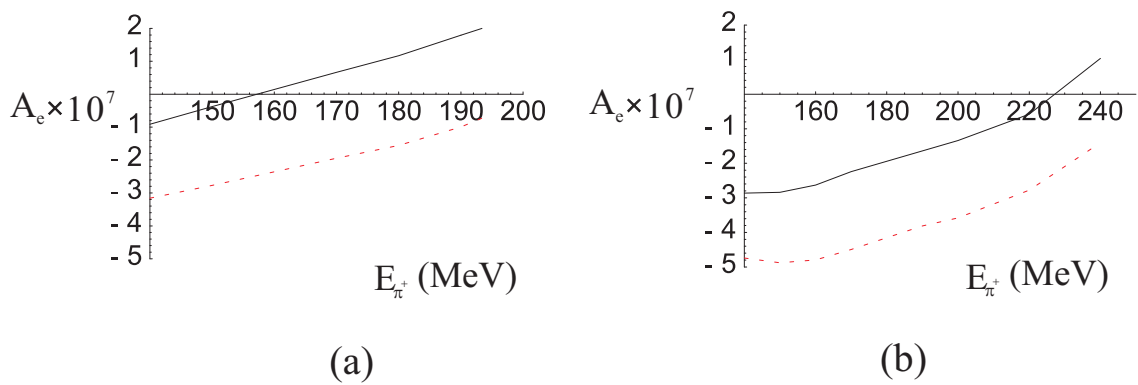}}
\end{center}
\caption{The electron-helicity asymmetry $A_e$ as a function
of the pion energy at (a) $E_e=200$ and (b)
$250$ MeV. The curves have the same significance as in Fig. 5.}
\label{fig:asym}
\end{figure}

We have also shown the asymmetry as a function of the
pion energy in Fig. 6. When the incident electron
energy is 200 MeV, the total asymmetry is again small. 
This is particularly true for most of the pions 
that are produced with 160 MeV energy. 
When the electron energy is 250 MeV, 
the cancelation between the Z-exchange and 
hadronic parity-violation is not complete. For
a large fraction of pions which are produced with 
energy between 160 and 180 MeV, the 
asymmetry is significant. Therefore, the
best way to extract $h_{\pi NN}^{(1)}$ from 
$ep\rightarrow e\pi^+ n$ is to measure pion production
asymmetry in the forward direction with a polarized 
electron beam of 250 MeV. 

For the incident electron energy of 250 MeV, 
the virtual photon energy is around 200 MeV (see fig. 2(a)). According
to our previous study \cite{CJ}, the unpolarized cross section
at this energy can be described well by the 
LO HB$\chi$PT calculation. Therefore, the applicability
of chiral perturbation theory in studying the parity
violating effects in the electroproduction
is qualitatively similar to that in photoproduction.
In particular, we believe that the higher-order
corrections are ${\cal O}(\epsilon/m_N)$, where
$\epsilon$ stands for $m_\pi$, $\omega$, $\omega_\pi$, and 
$\Delta$.

To summarize, we have studied the possibility of 
measuring the parity-violating coupling $h_{\pi NN}^{(1)}$
from semi-inclusive $\pi^+$ electroproduction on the
proton target in the threshold region. For the incident
energy of 250 MeV, the electron single-spin asymmetry
is estimated around $-2\times 10^{-7}$ in the forward 
direction. This asymmetry can be measured to an accuracy of
50\% (this implies an accuracy of $2 \times 10^{-7}$ in $h_{\pi
NN}^{(1)}$) with a
luminosity of $10^{38}$ sec$^{-1} \cdot$
cm$^{-2}$ and a running time of $10^7$ sec. 
A more careful study of the higher-order effects will 
be published later \cite{CJ2}.

\acknowledgements

We thank D. Beck, E. Beise, E.M. Henley, R. McKeown, R. Suleiman, and
B. Wojtsekhowski 
for discussions on the experimental issues. We also thank C. Jung for
help with the numerical calculations.
This work is supported in part by the U.S. Dept. of Energy
under grant No. DE-FG02-93ER-40762.

\end{document}